\documentclass{ws-p8-50x6-00}

\begin{document}

\title{A New Renormalization Group for Hamiltonian Field Theory}

\author{Robert J. Perry and  S{\'e}rgio Szpigel}

\address{Physics Department, The Ohio State University, Columbus, OH 43210,
USA\\E-mail: perry.6@osu.edu, szpigel@pacific.mps.ohio-state.edu}


\maketitle

\abstracts{
The Schr{\"o}dinger equation with a two-dimensional $\delta$-function
potential is a simple example of an asymptotically free theory that
undergoes dimensional transmutation. Renormalization requires the
introduction of a mass scale, which can be lowered perturbatively until
an infrared cutoff produced by non-perturbative effects such as
bound state formation is encountered. We outline the effective field
theory and similarity renormalization group techniques for producing
renormalized cutoff hamiltonians, and illustrate the control of
logarithmic and inverse-power-law errors both techniques provide.}

\section{Motivation and Outline}

This article follows a more complete discussion in a longer article being
prepared for publication.\cite{sergio_rg} Elsewhere in these
Proceedings the possibility of deriving a valid constituent approximation in
light-front QCD is discussed.\cite{japan_lfqcd} A simple procedure that
produces a renormalized light-front QCD hamiltonian starts with the computation
of a cutoff hamiltonian using a similarity renormalization
group (SRG) \cite{wilgla,wegner,glazek,walhout} and coupling
coherence.\cite{oehme,coupco,brent} In this article we illustrate this step
using the two-dimensional $\delta$-function in a one-body Schr{\"o}dinger
equation.\cite{jackiw} We also discuss a related approach,
effective field theory (EFT).\cite{Lepage,EFT}

Locality is the source of perturbative divergences in field theory, and it
produces similar effects in simple one-body quantum mechanics problems.
Locality leads us to consider potentials consisting of
$\delta$-functions and their derivatives. These potentials produce divergences
that require the introduction of a mass scale. Changing this mass scale should
be equivalent to changing a dimensionless coupling constant ({\it i.e.},
dimensional transmutation) for a scale-invariant local hamiltonian. The
two-dimensional $\delta$-function problem is solved in the second section,
providing the {\it data} for our error analysis of EFT and SRG calculations.

Problems requiring renormalization result from an arbitrarily large number of
degrees of freedom being coupled. This is clear in momentum space, where the
off-diagonal matrix elements of a $\delta$-function are constants. Divergences
result from an infinite number of scales being directly coupled, and
renormalization requires a cutoff $\lambda$ to regulate the divergences.
Physical results such as binding energies should be independent of $\lambda$,
so there should exist a line of cutoff hamiltonians $H_\lambda$ that all
produce the same physical results. A renormalization group (RG) is built from
transformations, ${\cal T}(\lambda,\lambda')$, that change the cutoff. ${\cal
T}$ connects hamiltonians at different scales and in the SRG ${\cal T}$ is
typically a unitary transformation.

In EFT ${\cal T}$ exists in principle, but $H_\lambda$ is determined at each
scale by constraining the couplings in front of a finite number of local
operators to fit data. A simple scale transformation leads to the classification
of these operators as relevant (powers of the cutoff appear), marginal (the
cutoff appears only in logarithms that are absorbed in the running marginal
coupling), and irrelevant (inverse powers of the cutoff appear). The same
operators arise in the SRG, where their strengths are approximated by
expansions in powers of a running marginal coupling. For the simple
two-dimensional $\delta$-function there are no relevant operators and only one
marginal operator. There are always an infinite number of irrelevant operators,
and these become increasingly important as the cutoff approaches a
non-perturbative energy scale in the problem.

The easiest way to clarify these points is by example. We show a complete
solution of the two-dimensional $\delta$-function problem, introducing the
K-matrix to compute scattering phase shifts. There is one bound state, and its
binding energy sets the scale for non-perturbative physics. The EFT hamiltonian
is produced by truncating the series of irrelevant operators, and then fitting
a finite number of remaining couplings to an equal number of low-energy
scattering constraints. This leads to power-law suppression of errors in the
predicted binding energy, with the expansion clearly breaking down as the cutoff
approaches the binding energy. The SRG hamiltonian is produced by solving a
differential equation for $H_\lambda$ subject to the constraint that all
irrelevant couplings are analytic functions of the single marginal coupling.

\section{The two-dimensional $\delta$-function potential}

Consider the Schr{\"o}dinger equation in two dimensions with an
attractive Dirac $\delta$-function potential:\cite{jackiw}

\begin{equation}
-{\bf \nabla}_{\bf r}^2 \Psi({\bf r})-\alpha_0 \;\delta^{(2)}({\bf r}) \;
\Psi({\bf r})=E \; \Psi({\bf r})\; .
\end{equation}
\noindent
The coupling  $\alpha_0$ is dimensionless, so the hamiltonian is scale
invariant (i.e., there is no intrinsic length or energy scale). It is
relatively straightforward to solve this equation in position space by
introducing a convenient distribution function to regulate the
$\delta$-function, but we will work in momentum space to stay close to the
EFT and SRG calculations.

The Schr\"{o}dinger equation in momentum space is,
\begin{equation}
p^2 \;  \Phi({\bf p})-\frac{\alpha_0}{(2\pi)^2}\; \int \; d^{2}q
\;\;\Phi({\bf q})=E \; \Phi({\bf p}) \; .
\label{semom2}
\end{equation}

\noindent
As a consequence of scale invariance, if there is a bound state solution to
Eq.(\ref{semom2}) then it will admit solutions for any $E<0$. This 
corresponds to a continuum of bound states with energies extending down
to $-\infty$, so the system is not bounded from below. By rearranging the
terms in the Schr\"{o}dinger equation we obtain   
\begin{equation}
\Phi({\bf p})=\frac{\alpha_0}{2\pi}\; \frac{\Psi(0)}{(p^2+E_0)} \; ,
\label{2ds}
\end{equation}
\noindent
where $\Psi(0)$ is the position space wave-function at the origin and
$E_0 =-E >0$ is the binding energy.

To obtain the eigenvalue condition for the binding energy, we can
integrate both sides of Eq.~(\ref{2ds}):
\begin{equation}
1=\frac{\alpha_0}{2\pi}\;  \int_{0}^{\infty} \; dp \; p \;
\frac{1}{(p^2+E_0)} \; .
\label{bsi2}
\end{equation}
\noindent
The integral on the r.h.s. diverges logarithmically, so the problem is
ill-defined. 

Renormalization requires that we regulate the divergence and absorb cutoff
dependence in the coupling. First, we regulate the integral with a momentum
cutoff, obtaining
\begin{equation}
1=\frac{\alpha_0}{2\pi}\;  \int_{0}^{\Lambda} \; dp \; p \;
\frac{1}{(p^2+E_0)}=\frac{\alpha_0}{4\pi}\; {\rm
ln}\left(1+\frac{\Lambda^2}{E_0}\right) \; ,
\end{equation}
\noindent
so that 
\begin{equation}
E_0=  \frac{\Lambda^2}{e^{-\frac{4\pi}{\alpha_0}}-1} \; .
\end{equation}

Clearly, if the coupling $\alpha_0$ is fixed then $E_0 \rightarrow
\infty$ as $\Lambda \rightarrow \infty$. In order to eliminate the
divergence and produce a finite, well-defined bound state we can
renormalize the theory by demanding that the coupling runs with the
cutoff $\Lambda$ in such a way that the binding energy remains fixed as
the cutoff is removed:
\begin{equation}
\alpha_0 \rightarrow \alpha_{\Lambda}=\frac{4\pi}{{\rm
ln}\left(1+\frac{\Lambda^2}{E_0}\right)} \; .
\label{couprun}
\end{equation}

The dimensionless renormalized running coupling $\alpha_{\Lambda}$  that
characterizes the strength of the interaction is therefore replaced by a
new (dimensionful) parameter $E_0$, the binding energy of the system.
This is a simple example of dimensional transmutation: even though the
original ``bare'' hamiltonian is scale invariant, renormalization
leads to a scale that characterizes the physical observables.
Note that $E_0$ can be chosen arbitrarily, fixing the energy scale of the
underlying (renormalized) theory. It is also interesting to note that the
renormalized running coupling $\alpha_{\Lambda}$ vanishes as $\Lambda
\rightarrow \infty$ and so the theory is asymptotically free.

This renormalized hamiltonian can be used to compute other observables.
The usual prescription for the calculations is to obtain the solutions
with the cutoff in place and then let it approach $\infty$, using the running
coupling in Eq. {\ref{couprun}}. If an exact calculation can be implemented, the
final results should be independent of the regularization and renormalization
schemes. As an example, we calculate the scattering wave function,
\begin{equation}
\Phi_{k}({\bf p})=\delta^{(2)}({\bf p}-{\bf k})+\frac{\alpha_{\Lambda}}{2
\pi}\; \frac{\Psi(0)}{(p^2-k^2+i \; \epsilon)} \; ,
\end{equation}
\noindent
where $k=\sqrt{E}$.
\noindent
Integrating both sides over ${\bf p}$ with a cutoff $\Lambda$ in place,
we obtain
\begin{equation}
\Psi(0)=\frac{1}{2\pi} \; \left[1-\frac{\alpha_{\Lambda}}{4\pi}\; {\rm
ln}\left(1+\frac{\Lambda^2}{-k^2+i\epsilon}\right)\right]^{-1} \; ;
\end{equation}
\noindent
thus,
\begin{equation}
\alpha_{\Lambda} \; \Psi(0)=\frac{1}{2\pi} \; \left[\frac{1}{4\pi}\;{\rm
ln}\left(1+\frac{\Lambda^2}{E_0}\right)-\frac{1}{4\pi}\; {\rm
ln}\left(1+\frac{\Lambda^2}{-k^2+i\epsilon}\right)\right]^{-1} \; .
\end{equation}
\noindent
In the limit $\Lambda \rightarrow \infty$ we obtain:
\begin{equation}
\alpha_{\Lambda} \; \Psi(0)=\frac{2}{{\rm
ln}\left(\frac{k^2}{E_0}\right)-i \; \pi} \; .
\end{equation}
\noindent
The resulting scattering wavefunction is then given by
\begin{equation}
\Phi_{k}({\bf p})=\delta^{(2)}({\bf p}-{\bf k})+\frac{1}{2 \pi}\;
\frac{2}{(p^2-k^2)}\left[{\rm ln}\left(\frac{k^2}{E_0}\right)-i \;
\pi\right]^{-1} \; .
\end{equation}
\noindent
Only S-wave scattering occurs, because the centrifugal
barrier completely screens the $\delta$-function potential for higher angular
momentum.

For our numerical
calculations we used the K-matrix, which has mixed boundary conditions that
lead to a principal value. The K-matrix
Lippmann-Schwinger equation with the renormalized potential is given by:
\begin{equation}
K({\bf p},{\bf p'};k)=V({\bf p},{\bf p'})+{\cal P}\int \; d^{2}q \; \;
\frac{V({\bf p},{\bf q})}{k^2-q^2} \; K({\bf q},{\bf p'};k) \;
.
\end{equation}
\noindent
Since only S-wave scattering takes place we can integrate over the angular
variable, obtaining 
\begin{equation}
K^{(\rm l=0)}( p,p';k)=V^{(\rm l=0)}(p,p')+{\cal P}\int_{0}^{\Lambda} \; dq \; q
\;  \frac{V^{(\rm l=0)}(p,q)}{k^2-q^2} \; K^{(\rm l=0)}(q,p';k) \; ,
\end{equation}
\noindent
where
\begin{equation}
V^{(\rm l=0)}(p,p')=-\frac{\alpha_{\Lambda}}{2\pi} \; .
\end{equation}

The Lippmann-Schwinger equation for the ``on-shell" K-matrix is given by:
\begin{eqnarray}
K^{(\rm l=0)}(k)=-\frac{\alpha_{\Lambda}}{2\pi}-\frac{\alpha_{\Lambda}}{2\pi}\;
K^{(\rm l=0)}(k) \; {\cal P}\int_{0}^{\Lambda} \; dq \; q \;  \frac{1}{k^2-q^2}.
\end{eqnarray}
\noindent
Solving this equation and taking the limit $\Lambda \rightarrow \infty$,
we obtain the exact on-shell K-matrix:
\begin{equation}
K_0(k)= -\frac{2}{{\rm ln}\left(\frac{k^2}{E_0}\right)}\; .
\end{equation}
\noindent
Using
\begin{equation}
k \;{\rm cot}\; \delta_0(k)=-\frac{2 \; k}{\pi}\frac{1}{K_0(k)} \; ,
\end{equation}
\noindent
we can obtain the exact phase-shifts:
\begin{equation}
{\rm cot} \; \delta_{0}=\frac{1}{\pi}\; {\rm ln}\left(\frac{k^2}{E_0}\right)\; .
\end{equation}

\section{Effective field theory}

\subsection{EFT formalism}

EFT provides a systematic procedure for improving the approximate description 
of composite effective degrees of freedom and their dynamics by increasing the 
number of local effective interactions. The effects of high-energy  degrees of
freedom are incorporated in effective interactions  whose strength is adjusted
to fit appropriate low-energy results. The keys to maintaining predictive power
in EFT are approximate  locality and an expansion in powers of a small ratio of
low-to-high energy  scales or equivalently, short-to-long distance scales.

We follow Lepage's treatment of EFT.\cite{Lepage}  For simplicity we use
a separable hamiltonian with a cutoff $\lambda$ that smoothly regulates
interactions involving high energy states:
\begin{eqnarray}
V_{\lambda}(p,p')&=&\left[C_0(\lambda) + C_2(\lambda) \; 
\frac{(p^2+p'^2)}{2\lambda^2}+C_4(\lambda) \; 
\frac{(p^4+p'^4)}{4\lambda^4}+C'_4(\lambda) \; \frac{p^2 \; p'^2}{2\lambda^4}+ 
\dots \right] \nonumber\\
&\times& e^{-\frac{p^2}{2\lambda^2}}\; e^{-\frac{p'^2}{2\lambda^2}} \; .
\label{pe}
\end{eqnarray}
\noindent
The parameters $C_i$ in the expansion can be determined by  fitting data for
low-energy processes. Here we  use as {\it data} the values for the inverse
on-shell K-matrix, calculated with the exact theory. We follow the method
described by Steele and Furnstahl \cite{furnste} with the parameters fixed so
that EFT gives the same momentum expansion for the inverse on-shell K-matrix as
the exact theory to a given order.

The Lippmann-Schwinger equation for the K-matrix with is
\begin{equation}
K_{\lambda}(p,p';k)=V_{\lambda}(p,p')+{\cal P}\int \; d^{2}q \; \; 
\frac{V_{\lambda}(p,q)}{k^2-q^2} \; K_{\lambda}(q,p';k) \; ,
\end{equation}
\noindent
where ${\cal P}$ denotes the principal value. The equation refers to the
S-wave potential (angular variable integrated out) and $d^{2}q=q \; dq$. To
obtain the K-matrix  we solve the Lippmann-Schwinger equation
nonperturbatively.\cite{cohen,birse,geg} The effective potential given in
Eq.(\ref{pe}) (truncated  at order  $p^4$) can be written in the form
\begin{equation}
V_{\lambda}(p,p')=e^{-\frac{p^2}{2\lambda^2}} \; 
e^{-\frac{p'^2}{2\lambda^2}}\sum_{i,j=0}^{2} p^{2i}\; \Lambda_{ij}\; p'^{2j} \; 
,
\end{equation}
\noindent
where $\Lambda_{ij}$ are the matrix elements of
\begin{equation}
{\bf \Lambda}=\left( \begin{array}{ccc} 
C_0(\lambda) & C_2(\lambda)/2\lambda^2  & C_4(\lambda)/4\lambda^4\\ \\
C_2(\lambda)/2\lambda^2  & C'_4(\lambda)/2\lambda^4&0\\ \\ 
C_4(\lambda)/4\lambda^4&0&0 
\end{array} \right) \; .
\end{equation}
\noindent
The solution of the Lippmann-Schwinger equation can then be written in the form

\begin{equation}
K_{\lambda}(p',p;k)=e^{-\frac{p^2}{2\lambda^2}} \; 
e^{-\frac{p'^2}{2\lambda^2}}\sum_{i,j=0}^1 p^{2i}\; \tau_{ij}(k)\; p'^{2j} \; .
\label{Kmat}
\end{equation}
\noindent
The unknown matrix ${\bf \tau}(k)$ satisfies the equation
\begin{equation}
{\bf \tau(k)}={\bf \Lambda} + {\bf \Lambda} \;{\bf {\cal I}}(k) \;{\bf 
\tau(k)}\; ,
\label{tau}
\end{equation}
\noindent
where
\begin{equation}
{\bf {\cal I}}=\left( \begin{array}{ccc} 
I_0(k) & I_1(k)  & I_2(k)\\ \\
I_1(k) & I_2(k)  & I_3(k)\\ \\ 
I_2(k) & I_3(k)  & I_4(k)
\end{array} \right) \; ,
\end{equation}

\noindent
with
\begin{equation}
I_{n}(k)= {\cal P}\int dq \; q^{2n+D-1}\;  \frac{1}{k^2 - q^2}\; 
e^{-\frac{q^2}{\lambda^2}}\; . 
\end{equation}
\noindent
Solving Eq. (\ref{tau}) for ${\bf \tau}$ analytically  and substituting the 
solution in  Eq. (\ref{Kmat}) we obtain the effective K-matrix. 

It is straightforward to extend the calculation to higher orders. We fit
the difference between the effective and exact inverse on-shell K-matrix
($p^2=p'^2=k^2$) to an interpolating polynomial in  $k^2/{\lambda^2}$  to
the highest possible order,
\begin{equation}
\Delta \; \left[\frac{1}{K}\right]=A_0+A_2 \;  \frac{k^2}{\lambda^2}+A_4 \; 
\frac{k^4}{\lambda^4}+ \cdots \; .
\label{del}
\end{equation}
\noindent
The coefficients $A_i$ are minimized with respect to the variation in the 
parameters $C_i$ of the effective potential. The number of coefficients that 
can be minimized is given by the number of parameters appearing in the 
effective potential. 

By adjusting only the parameter $C_0(\lambda)$ we should eliminate the leading 
error. As each term is added to the effective potential, followed by the 
adjustment of the respective parameter, we expect the errors in the 
on-shell K-matrix to be systematically reduced by {\it powers of 
$k^2/{\lambda^2}$}. Thus, in a log-log plot for $\Delta[1/K]$ we expect to 
obtain straight lines with slope given by the dominant power of $k/{\lambda}$ 
in the error.

The expansion given by Eq.( \ref{del}) becomes invalid at best when the 
momenta involved are of the same order as the cutoff, where one expects the 
short distance effects to be directly resolved. This point corresponds to the 
radius of convergence of the effective theory. Using the effective potential
(truncated at a given order) with the parameters fixed by fitting the inverse
on-shell K-matrix  we diagonalize the  effective hamiltonian numerically,
obtaining a bound-state energy that can be compared to the exact result.

\subsection{EFT results for the two-dimension $\delta$-function}

The EFT power counting scheme is simple. For the leading-order prediction, 
we use the effective potential with one parameter. Evaluating the difference
$\Delta \; [1/K^{(0)}]=1/K_{\lambda}^{(0)}-1/K_0$ and expanding it in powers of 
$k/{\lambda}$ we obtain:
\begin{equation}
\Delta \; \left[\frac{1}{K^{(0)}}\right]=
\left[\frac{1}{C_0}-\frac{\gamma}{2}+\frac{1}{2}\; {\rm 
ln}\left(\frac{\lambda^2}{E_0}\right)\right]
+\left[\frac{1}{C_0}-\frac{1}{2}\right]\; \frac{k^2}{\lambda^2}+{\cal 
O}\left(\frac{k^4}{\lambda^4}\right)\; .
\end{equation}

By choosing 
\begin{equation}
C_0(\lambda)=\frac{-2}{{\rm ln}\left(\frac{\lambda^2}{E_0}\right)-\gamma}
 \; ,
\end{equation}
\noindent
the leading logarithmic error is eliminated and the remaining errors are 
dominated by the term 
\begin{equation}
{\cal O}\left(\frac{k^2}{\lambda^2}\right)
=\frac{1}{2}\left[\gamma-{\rm 
ln}\left(\frac{\lambda^2}{E_0}\right)-1\right]\frac{k^2}{\lambda^2}\; .
\end{equation}

To evaluate the errors in the binding energy we use the effective potential with
one, two and four parameters adjusted to fit the inverse K-matrix. We solve the
Schr\"{o}dinger equation numerically for different values of the cutoff
$\lambda$. In Fig. 1 we show in a log-log plot the absolute values for the
relative errors in the bound-state energy as a function of $E_0/{\lambda^2}$.
As expected, we see straight lines with slope given by the  dominant power of
$E_0/{\lambda^2}$ in the error. Note that the dominant logarithmic errors are
completely absorbed by adjusting the parameter $C_0$ and when more parameters
are adjusted we obtain power-law improvement.

\begin{figure}
\epsfxsize=24pc
\center{\epsfbox{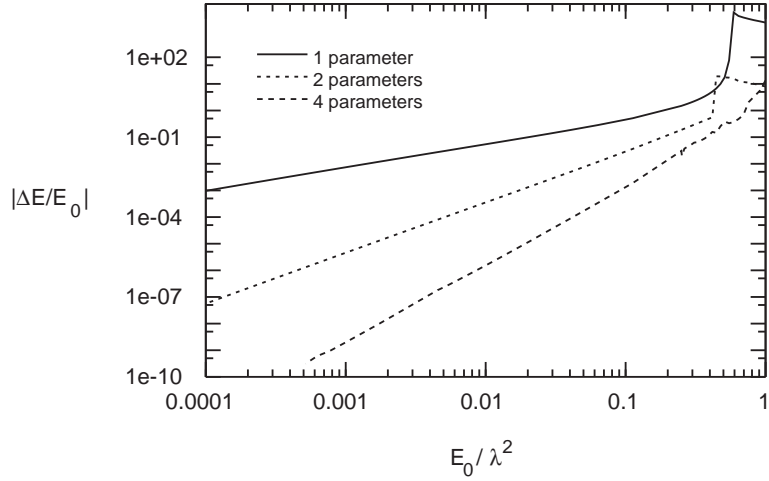}}
\caption{The EFT error in the binding energy for the two-dimensional 
delta-function using one, two and four parameters. The exact theory is fixed by 
choosing $E_0=1$.}
\end{figure}

\section{The similarity renormalization group}

\subsection{SRG formalism}

The similarity renormalization group \cite{wilgla,wegner} is made of
similarity transformations that run a smooth cutoff on the magnitude of
off-diagonal matrix elements. As the cutoff is lowered, the hamiltonian is
forced toward the diagonal. It is straightforward to show that an infinitesimal
transformation produces a differential equation governing the evolution of the
hamiltonian. G{\l}azek and Wilson considered very general transformations in
their original work,\cite{wilgla} but we use the simple transformation
developed by Wegner.\cite{wegner} The infinitesimal transformation is
\begin{equation}
\frac{d H_s}{ds}=[H_s,[H_s,h]
\;.
\end{equation}
\noindent
Here $H_s=h+V_s$, where $h$ is the free hamiltonian, evolves with a flow
parameter $s$ that ranges  from $0$ to $\infty$. The flow-parameter has
dimensions $1/({\rm energy})^2$  and can be expressed in terms of the similarity
cutoff $\lambda$ by the  relation $s=1/\lambda^2$.  Wegner advocates using the
full diagonal part of the hamiltonian instead of using $h$ in the
transformation.

The reduced interaction, ${\overline V}_{sij}$ (the interaction with a 
gaussian similarity cutoff factored out) is defined by
\begin{equation}
V_{sij}= e^{-s\Delta_{ij}^2} \; {\overline V}_{sij}
\;,
\end{equation}
\noindent where $\Delta_{ij}=p_i^2-p_j^2$ for our simple transformation.
\noindent The gaussian factor forces the interaction toward the diagonal to
leading order, and the equation for the evolution of ${\overline V}$ starts at
second order.

The flow equation for the reduced interaction is
\begin{equation}
\frac{d{\overline V}_{sij}}{ds}=\sum_k\left(\Delta_{ik}+\Delta_{jk}\right)
\; {\overline V}_{sik} \; {\overline V}_{skj} \; e^{-2s
\Delta_{ik}\Delta_{jk}}
\;,\label{eq:wegnerbigone}
\end{equation}
\noindent
where we use 
$\Delta_{ij}^2-\Delta_{ik}^2-\Delta_{jk}^2=-2\Delta_{ik}\Delta_{jk}$.

To solve this equation we impose a 
boundary condition, $H_s |_{_{s \rightarrow s_0}} \equiv H_{s_0}$, where $s_0$ 
can be expressed in terms of the large similarity cutoff  as $s_0=1/\Lambda^2$. 
Next we make a perturbative expansion,
\begin{equation}
{\overline V}_{s}=
{\overline V}_{s}^{^{(1)}}
+{\overline V}_{s}^{^{(2)}}+\cdots\;,
\end{equation}
where the superscript implies the order in the running canonical coupling. All
counterterms are determined using coupling coherence, which assumes ${\overline
V}$ is a perturbative function of $\alpha$.

At first order we have
\begin{equation}
\frac{d{\overline V}_{s_{ij}}^{^{(1)}}}{ds}=0\;,
\end{equation} 
which implies
\begin{equation}
{\overline V}_{sij}^{^{(1)}}={\overline V}_{s_{_{0}}ij}
\;,
\end{equation}
where $s$ is the final scale. Because of the dimensions of the flow parameter 
we have $s > s_{_{0}}$, corresponding to a smaller cutoff.  
The ``no cutoff limit" corresponds to $s_{_{0}}\longrightarrow 0$.

At second order we have
\begin{equation}
\frac{d{\overline 
V}_{sij}^{^{(2)}}}{ds}=\sum_k\left(\Delta_{ik}+\Delta_{jk}\right)
\; {\overline V}_{s_{_{0}}ik}\; {\overline V}_{s_{_{0}}kj} \; e^{-2s
\Delta_{ik}\Delta_{jk}}
\;.
\end{equation}
Integrating, we obtain
\begin{eqnarray}
{\overline V}_{sij}^{^{(2)}}&=&\frac{1}{2}\sum_k
{\overline V}_{s_{_{0}}ik}\; {\overline V}_{s_{_{0}}kj} \; 
\left(\frac{1}{\Delta_{ik}}+\frac{1}{\Delta_{jk}}\right)\times\nonumber\\
&&~~~~~~~~~~\times \left[e^{-2 s_0 
\Delta_{ik}\Delta_{jk}}-e^{-2s\Delta_{ik}\Delta_{jk}}\right]
\;.
\end{eqnarray}

The renormalized  hamiltonian for the non-relativistic delta-function potential
is given by 
\begin{equation}
H_{\lambda}({\bf p},{\bf p'})=p^2 \delta^{(2)}({\bf p}-{\bf p'}) + 
e^{-\frac{(p^2-p'^2)^2}{\lambda^4}}\; \left[{\bar V}_{\lambda}^{(1)}({\bf 
p},{\bf p'})+{\bar V}_{\lambda}^{(2)}({\bf p},{\bf p'})+... \right]\; ,
\label{renh}
\end{equation}
\noindent
where
\begin{eqnarray}
{\bar V}_{\lambda}^{(1)}({\bf p},{\bf 
p'})&=&-\frac{\alpha_{\lambda,i}}{(2\pi)^2} \; ,\\
{\bar V}_{\lambda}^{(2)}({\bf p},{\bf p'})&=&\alpha_{\lambda,i}^2 \; 
F^{(2)}_{s}({\bf p},{\bf p'}) \; , \\
{\bar V}_{\lambda}^{(n)}({\bf p},{\bf p'})&=&\alpha_{\lambda,i}^n \; 
F^{(n)}_{s}({\bf p},{\bf p'}) \; .
\end{eqnarray}
\noindent
Here $\lambda$ is a momentum cutoff related to the flow parameter by 
$s=1/{\lambda^4}$ and $i$ denotes the order of the calculation for the 
running coupling.

The renormalized hamiltonian can be used to compute eigenvalues and 
eigenstates. Since the hamiltonian is derived perturbatively we expect cutoff 
dependent errors in the observables. Formally, we can regroup the terms in the 
renormalized hamiltonian and write it as a momentum  expansion similar to the 
one for the EFT in Eq.~(\ref{pe}). The difference is that the expansion 
parameters are perturbative functions of the running coupling
$\alpha_{\lambda}$. 

There are two interdependent sources of errors in the perturbative 
similarity renormalization group. First, $\alpha_\lambda$ is approximated and
this leads to inverse-logarithmic errors. Second, the coefficients of all
irrelevant operators are approximated perturbatively, and this leads to
products of inverse-logarithmic and inverse-power-law errors.

In our examples we focus on the bound state errors. We fix the coupling at
one scale and use the flow-equation to  obtain the coupling as a function of the
cutoff $\lambda$ to a given order. We then perform a sequence of bound-state
calculations with better approximations for the hamiltonian. Once the sources
of errors are identified, it becomes  relatively simple to analyze
order-by-order how such errors scale with $\lambda$. In principle,
to completely eliminate the errors proportional to some power $m$ in the
momentum expansion we should use the similarity  hamiltonian with the exact
running coupling (renormalized to all orders) and  include the contributions up
to ${\cal O}(p^m/{\lambda^m})$ coming from all  effective interactions (all
orders in $\alpha_{\lambda}$). We should emphasize that in a realistic
calculation we would fit the coupling $\alpha_{\lambda}$ to an observable. This
nonperturbative renormalization  eliminates the dominant source of errors we
display in SRG calculations in this  article. We choose to renormalize the
coupling perturbatively here because the only observable we compute
is the single bound state energy of a  delta-function potential, and fitting
this energy would prevent us from  displaying errors.

\subsection{SRG results for two-dimensional $\delta$-function}

G{\l}azek and Wilson have studied a discretized analog of the two-dimensional
$\delta$ function, which allowed them to perform both perturbative and
nonperturbative transformation.\cite{glazek} Their perturbative results are
similar to ours. In the two-dimensional case the canonical hamiltonian in
momentum space with a  delta-function potential can be written as 
\begin{equation}
H({\bf p},{\bf p'})=h({\bf p},{\bf p'})+V({\bf p},{\bf p'}) \; , 
\end{equation}
\noindent
where $h({\bf p},{\bf p'})=p^2 \delta^{(2)}({\bf p}-{\bf p'})$ corresponds to 
the free hamiltonian and $V({\bf p},{\bf p'})=-{\alpha_0}/(2\pi)^2$ corresponds 
to the Fourier transform of the delta-function potential.

Integrating out the angular variable, the flow equation obtained with Wegner's 
transformation in terms of matrix elements in the basis of free states is given 
by 
\begin{equation}
\frac{dV_s(p,p')}{ds}=-(p^2-p'^2)^2 \; V_{s}(p,p')-\int_{0}^{\infty}dk \; k \; 
(2 k^2-p^2-p'^2)\; V_{s}(p,k)\; V_{s}(k,p') \; .
\end{equation}
\noindent
In principle, we can set the boundary condition at $s=0$ (no cutoff), i.e, 
\begin{equation}
H_{s=0}(p,p')=H(p,p')=p^2 \delta^{(1)}(p-p')-\frac{\alpha_0}{2\pi} \; .
\end{equation}
\noindent
However, the hamiltonian with no cutoff produces logarithmic divergences. The
boundary condition must be  imposed at some other point, leading to dimensional
transmutation.
\noindent
The reduced interaction ${\bar V}_{s}(p,p')$ is defined such that
\begin{equation}
V_{s}(p,p')=e^{-s(p^2-p'^2)^2}\; {\bar V}_{s}(p,p') \; .
\end{equation}
\noindent
Assuming that $h$ is cutoff independent we obtain the flow equation for the 
reduced interaction, 
\begin{eqnarray}
\frac{d{\bar V}_{s}}{ds}=-e^{-2s\; p^2  p'^2}\; \int_{0}^{\infty}&&dk \; k \; 
(2 k^2-p^2-p'^2)\; e^{-2s[k^4-k^2(p^2+p'^2)]}\nonumber\\
&&\times {\bar V}_{s}(p,k)\; {\bar V}_{s}(k,p') \; .
\label{f2}
\end{eqnarray}

This equation is solved using  a perturbative expansion, starting with
\begin{equation}
{\bar V}^{(1)}_{s}(p,p')=-\frac{{\alpha}_s}{2\pi} \; .
\end{equation}
\noindent
We assume a coupling-coherent solution in the form of an expansion in powers of 
$\alpha_s/2\pi$, satisfying the constraint that the operators 
$F^{(n)}_{s}(p,p')$ vanish when $p=p'=0$,
\begin{equation}
{\bar 
V}_{s}(p,p')=-\frac{\alpha_s}{2\pi}+\sum_{n=2}^{\infty}\left(\frac{\alpha_{s}}{
2\pi}\right)^{n}\; F^{(n)}_{s}(p,p') \; .
\label{coc2}
\end{equation}
\noindent
Note that the expansion parameter is $\alpha_s/2\pi$.

Using  the solution  Eq.(\ref{coc2}) in Eq.(\ref{f2}) we obtain
\begin{eqnarray}
\frac{d{\bar V}_{s}}{ds}&=&-\frac{1}{(2\pi)^2}\; 
\frac{d{\alpha}_s}{ds}+\sum_{n=2}^{\infty}\frac{1}{(2\pi)^n}\left[n \; 
\alpha_{s}^{n-1}\; \frac{d{\alpha}_s}{ds}\; F^{(n)}_{s}(p,p')+\alpha_{s}^{n}\; 
\frac{dF^{(n)}_{s}(p,p')}{ds}\right]\nonumber\\
&=&\int_{0}^{\infty}dk \; k \; (2 k^2-p^2-p'^2)\; e^{-2s[p^2 
p'^2+k^4-k^2(p^2+p'^2)]}\nonumber\\
&\times&\left[-\frac{\alpha_s}{2\pi}+\sum_{n=2}^{\infty}\left(\frac{\alpha_{s}}
{2\pi}\right)^{n}\; 
F^{(n)}_{s}(p,k)\right]\left[-\frac{\alpha_s}{2\pi}+\sum_{m=2}^{\infty}
\left(\frac{\alpha_{s}}{2\pi}\right)^{m}\;
F^{(m)}_{s}(k,p')\right]
\end{eqnarray}
\noindent
This equation is solved iteratively order-by-order in $\alpha_s/2\pi$. Again, 
if $\alpha_s/2\pi$ is small the operator ${\bar V}^{(1)}_{s}(p,p')$ can be 
identified as the dominant term in the expansion of ${\bar V}_{s}(p,p')$ in 
powers of $p$ and $p'$. This operator corresponds to a 
marginal operator (since the coupling is dimensionless and there is no implicit 
mass scale). The higher-order terms correspond to irrelevant operators. 

For simplicity we use the interaction with a large cutoff on all momenta,
$\Lambda$, to define the {\it exact} theory. We define
\begin{equation}
\alpha_{s_0=0}=\alpha_{\Lambda}=\frac{4\pi}{{\rm 
ln}\left(1+\frac{\Lambda^2}{E_0}\right)} \; ,
\end{equation}   
\noindent
and set all irrelevant operators to zero at $s_0=0$. Note that the coupling 
$\alpha_{\lambda_0}$ is fixed at $\lambda_0=\infty$ by fitting the exact 
binding energy. With this definition the similarity hamiltonian with no cutoff 
becomes well-defined and we can set all of the similarity transformation 
boundary conditions at $s_0=0$. All integrals below are cut off ($p\ll \Lambda$)

At second-order we have
\begin{equation}
-\frac{1}{2 \pi}\; 
\frac{d{\alpha}_{s,\Lambda}}{ds}+\frac{1}{(2\pi)^2}\alpha_{s,\Lambda}^{2}\; 
\frac{dF^{(2)}_{s,\Lambda}(p,p')}{ds}=- \alpha_{s,\Lambda}^2 \; 
I_{s,\Lambda}^{(2)}(p,p')
\end{equation}
\noindent
where
\begin{eqnarray}
I_{s,\Lambda}^{(2)}(p,p')&=&\frac{1}{(2\pi)^2}\; \int_{0}^{\Lambda}dk \; k \; 
(2 k^2-p^2-p'^2)\; e^{-2s[p^2 p'^2+k^4-k^2(p^2+p'^2)]}\nonumber\\
&=&\frac{1}{(2\pi)^2}\; \frac{e^{-2s \; p^2 p'^2}}{4s} \; .
\end{eqnarray}
\noindent
The resulting second-order running coupling and irrelevant operator are given 
respectively by
\begin{equation}
\alpha_{s,\Lambda,2}=\frac{\alpha_{\Lambda}}{1-\frac{\alpha_{\Lambda}}{8\pi}\; 
\left[\gamma+{\rm ln}\left(2s \Lambda^4 \right)-{\rm Ei}\left(-2s \Lambda^4 
\right)\right]}
\end{equation}
\noindent
and
\begin{eqnarray}
F^{(2)}_{s,\Lambda}(p,p')&=&  \frac{1}{4}\left[\gamma+{\rm ln}(2s \; p^2 \; 
p'^2)-{\rm Ei}(-2s \; p^2 \; p'^2)\right]\nonumber\\
&+& \frac{1}{4}\left[\gamma+{\rm ln}(2s \Lambda^4)-{\rm Ei}(-2s 
\Lambda^4)\right]\nonumber\\
&-&\frac{1}{4}\left[\gamma+{\rm ln}\left(s \left[(p^2-\Lambda^2)^2 
+(p'^2-\Lambda^2)^2-(p^2-p'^2)^2\right]\right)\right.\nonumber\\
&&\left. \; \; \; \; \; -{\rm Ei}\left(-s \left[(p^2-\Lambda^2)^2 
+(p'^2-\Lambda^2)^2-(p^2-p'^2)^2\right]\right)\right]\; .
\end{eqnarray}

\begin{figure}
\epsfxsize=20pc
\center{\epsfbox{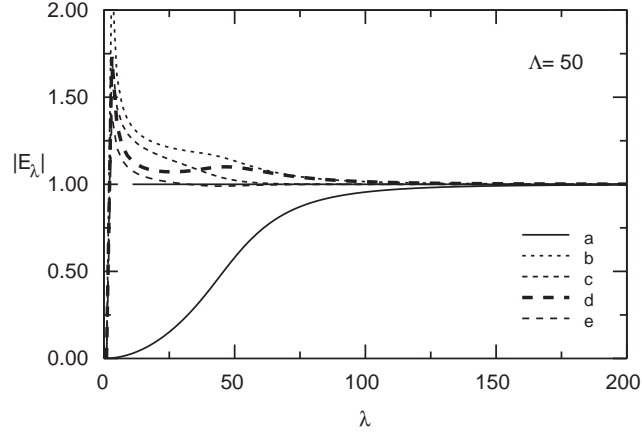}}
\caption{The SRG bound-state energy using various approximations for the
similarity hamiltonian. The exact theory is fixed by  choosing
$\Lambda=50$ and $E_0=1$.}
\end{figure}

\begin{figure}
\epsfxsize=24pc
\center{\epsfbox{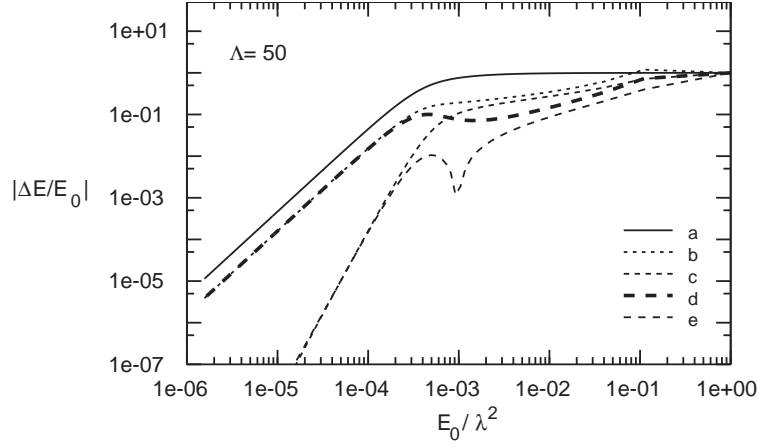}}
\caption{The SRG errors in the binding energy using various approximations for
the similarity hamiltonian. The exact theory is fixed by choosing $\Lambda=50$
and  $E_0=1$.}
\end{figure}

In Fig. 2 we show the binding-energy as a function of the cutoff and in Fig. 3
we show the errors in the binding-energy obtained using the following 
approximations for the potential with $\Lambda=50$:
\vspace{0.5cm}

\noindent
(a) marginal operator with coupling ($\alpha_{\lambda_0}$),
\begin{equation}
V_{\lambda,\Lambda}(p,p')=-\frac{\alpha_{\Lambda}}{2\pi}\; 
e^{-\frac{(p^2-p'^2)^2}{\lambda^4}} \; ;
\end{equation}

\noindent
(b) marginal operator with running coupling renormalized to second-order 
($\alpha_{\lambda,\Lambda,2}$),
\begin{equation}
V_{\lambda,\Lambda}(p,p')=-\frac{\alpha_{\lambda,\Lambda,2}}{2\pi}\; 
e^{-\frac{(p^2-p'^2)^2}{\lambda^4}} \; ;
\end{equation}

\noindent
(c) marginal operator plus second-order irrelevant operator with running 
coupling renormalized to second-order ($\alpha_{\lambda,\Lambda,2}, 
F^{(2)}_{\lambda}$),
\begin{equation}
V_{\lambda,\Lambda}(p,p')=\left[-\frac{\alpha_{\lambda,\Lambda,2}}{2\pi}+\left(
\frac{\alpha_{\lambda,\Lambda,2}}{2\pi}\right)^2 \; 
F^{(2)}_{\lambda,\Lambda}(p,p')\right] \; e^{-\frac{(p^2-p'^2)^2}{\lambda^4}} 
\; ;
\end{equation}

\noindent
(d) marginal operator with running coupling renormalized to third-order  
($\alpha_{\lambda,\Lambda,3}$),
\begin{equation}
V_{\lambda,\Lambda}(p,p')=-\frac{\alpha_{\lambda,\Lambda,3}}{2\pi}\; 
e^{-\frac{(p^2-p'^2)^2}{\lambda^4}} \; ;
\end{equation}

\noindent
(e) marginal operator plus second-order irrelevant operator with running 
coupling renormalized to third-order $\alpha_{\lambda,\Lambda,3}, 
F^{(2)}_{\lambda}$),
\begin{equation}
V_{\lambda,\Lambda}(p,p')=\left[-\frac{\alpha_{\lambda,\Lambda,3}}{2\pi} 
+\left(\frac{\alpha_{\lambda,\Lambda,3}}{2\pi}\right)^2 \; 
F^{(2)}_{\lambda,\Lambda}(p,p')\right] \; e^{-\frac{(p^2-p'^2)^2}{\lambda^4}}  
\; .
\end{equation}
\vspace{0.5cm}

In Fig. 2 we clearly see all of the binding energies approaching the exact
result, which we choose to be 1. If the coupling does not run (a) the energy
falls to zero as the cutoff is lowered. When the coupling is allowed to run to
second order (b) the binding energy deviates from the exact result more slowly
as the cutoff is lowered, and subsequent improved approximations further remove
dependence on $\lambda$.

Fig. 3 is a log-log plot of the relative error in the binding energy as a
function of $E_0/\lambda^2$. When $\lambda \gg \Lambda$ the errors scale like
inverse powers of $E_0/\lambda^2$. They are forced to approach the correct
result as $\lambda \rightarrow \infty$, and the exact running coupling deviates
from the approximate running coupling as a power of $\Lambda/\lambda$ in this
region, even when the approximate coupling is not allowed to run. The
coefficients of these powers are large, and the analysis changes as $\lambda
\rightarrow \Lambda$. The errors become logarithmic for (a), but they scale as
an inverse-logarithm squared in (b) where the coupling runs to second-order.
This removes the dominant inverse-logarithmic errors, and we can clearly see
further improvement when the coupling runs to third-order in (d). The addition
of irrelevant operators in (c) and (e) are important, but the third-order error
in the running coupling becomes as large as the error from the leading
irrelevant operators, which shows up when (d) crosses (c). When the
second-order irrelevant operators are added with the coupling running to third
order, we obtain approximate power-law improvement over (b).

\section{Conclusion}

Renormalization not only removes unphysical divergences, it also allows us to
work with finite cutoffs while controlling the errors they introduce.
Finite cutoffs are often required for nonperturbative calculations, and the
precise control of the errors they introduce is essential.

In light-front QCD we can use the SRG to compute the hamiltonian order-by-order
in the canonical QCD coupling, $\alpha_\lambda$. We are free to choose $\lambda$
so that multi-parton high-energy states effectively decouple from few-parton
low-energy states, as long as this happens before $\alpha_\lambda$ becomes too
large. We adjust the coupling non-perturbatively by fitting
low-energy data. This should remove all inverse-logarithmic errors, leaving
only powers of inverse-logarithms and inverse-powers of the cutoff. We can
systematically lower the errors by adding irrelevant operators, all of which
are fixed by coupling coherence.

\section*{Acknowledgments}
One of us (R.J.P.) is very grateful to the hosts of the 1998 YITP-Workshop on
QCD and Hadron Physics. The opportunity to see Kyoto and to interact with so
many excellent young Japanese physicists was fantastic. We would also like to
acknowledge many useful discussions with Brent Allen, Dick Furnstahl, Stan
G{\l}azek, Roger Kylin, Rick Mohr, Jim Steele, and Ken Wilson. This work was
supported by National Science Foundation grant PHY-9800964, and S.S. is a
CNPq-Brazil fellow (proc. 204790/88-3).

\end{document}